\documentclass[seceq,epsfig]{ptptex}
\usepackage{epsfig}

\title{
Tetra-Quark Resonances in Lattice QCD%
}

\author{
Hideo \textsc{Suganuma,}$^{1}$\footnote{ e-mail address:
suganuma@ruby.scphys.kyoto-u.ac.jp} 
Kyosuke \textsc{Tsumura}$^{1}$,
Noriyoshi \textsc{Ishii}$^{2}$ \\
and 
Fumiko~\textsc{Okiharu}$^{3}$
}

\inst{
$^{1}$ Department of Physics, Kyoto University, Kyoto 606-8502, Japan\\
$^{2}$ Center for Computational Science, Tsukuba University, 305-8577, Japan\\
$^{3}$ Department of Physics, Nihon University, Tokyo 101-8308, Japan
}

\abst{
We study $qq \bar q \bar q$-type four-quark (4Q) systems   
in SU(3)$_c$ anisotropic quenched lattice QCD,
using the $O(a)$-improved Wilson (clover) fermion at $\beta=5.75$ on $12^3 \times 96$ 
with renormalized anisotropy $a_s/a_t=4$.
For comparison, we first investigate the lowest $q\bar q$ scalar meson from 
the connected diagram and find its large mass of about 1.32GeV after chiral extrapolation, 
and thus the lowest $q\bar q$ scalar meson corresponds to $f_0(1370)$.
We investigate the lowest 4Q state 
in the spatially periodic boundary condition, and find that 
it is just a two-pion scattering state, as is expected. 
To examine spatially-localized 4Q resonances, 
we use the Hybrid Boundary Condition (HBC) method,  
where anti-periodic and periodic boundary conditions are imposed 
on quarks and antiquarks, respectively. 
By applying HBC on a finite-volume lattice, the threshold of the two-meson scattering state is 
raised up, while the mass of a compact 4Q resonance is almost unchanged.
In HBC, the lowest 4Q state appears slightly below the two-meson threshold.
To clarify the nature of the 4Q system, 
we apply the Maximum Entropy Method (MEM) for the 4Q correlator and 
obtain the spectral function of the 4Q system.
From the combination analysis of MEM with HBC, 
we finally conclude that the 4Q system appears as a two-pion scattering state 
and there is no spatially-localized 4Q resonance 
in the quark-mass region of $m_s< m_q <2m_s$.
}

\begin{document}

\maketitle

\section{Introduction}

There are five $0^{++}$ isoscalar mesons below 2GeV: 
$f_0$(400-1200), 
$f_0$(980), 
$f_0$(1370), 
$f_0$(1500) and 
$f_0$(1710). 
Among them, $f_0$(1370) is assigned to be the lowest $q\bar q$ scalar meson in the quark model \cite{PDG}, 
since the lowest $q\bar q$ scalar meson is $\ ^3P_0$ in the quark model 
and it turns to be rather heavy. 
$f_0$(1500) and $f_0$(1710) are expected to be the lowest scalar glueball \cite{PDG,ISM02} 
or an $s\bar s$ scalar meson \cite{PDG}.
Then, what are the two light scalar mesons, $f_0$(400-1200) and $f_0$(980)? 
This is the ``scalar meson puzzle", which is unsolved even at present. 
As a possible answer, in 1977, 
Jaffe proposed tetra-quark ($qq\bar q\bar q$) assignment \cite{J77}
for low-lying scalar mesons such as $f_0$(980) and $a_0$(980).

The tetra-quark is also interesting in terms of 
the recent experimental discoveries of charmed tetra-quark candidates, {\it e.g.}, 
D$_{\rm s0}^+$(2317), D$_{\rm s1}^+$(2460), X(3872) and Y(4260), 
which are expected to be tetra-quark states \cite{CP04,BG04}
from the consideration of their mass, narrow decay width and decay pattern.
In fact, simple quark-model assignment for D$_{\rm s0}^+$(2317) is $c\bar s(^3P_0)$,
but the P-wave $c\bar s$ state is much heavier than 2.32GeV \cite{BG04}.
Furthermore, there is a puzzle in the decay mode of D$_{\rm s0}^+$(2317): 
CLEO experiments show that 
the dominant decay mode is D$_{\rm s}^{+} \pi^0$ and 
its contribution is much larger than the radiative decay as 
$
\frac{\Gamma({\rm D}_{\rm s0}^+ \rightarrow {\rm D}_{\rm s}^{*+} \gamma)}
     {\Gamma({\rm D}_{\rm s0}^+ \rightarrow {\rm D}_{\rm s}^{+} \pi^0)} < 0.059.
$
If D$_{\rm s0}^+$(2317) is isoscalar such as $c\bar s$, 
this decay pattern implies that the main decay mode is isospin breaking process 
of $O(e^4)$ in QED, which is rather anomalous.
To explain this decay pattern, Terasaki proposed $I=1$ possibility of D$_{\rm s0}^+$(2317) \cite{T03}, 
which means the tetra-quark picture for D$_{\rm s0}^+$(2317) and predicts 
a doubly-charged meson D$_{\rm s0}^{++}$ 
as the isospin~partner. 

In this way, multi-quark systems ($qq\bar q \bar q$, $qqqq\bar q$, $\cdots$) have been studied 
as an interesting subject, and, recently, they are also investigated in lattice QCD 
\cite{OST05,IDS05,STIO06,TUOK05}
as well as ordinary $q\bar q$ and $qqq$ systems.\cite{TS0102,NNMS03,IDIST06}
In this paper, we investigate the four-quark (4Q) system and 
examine the scalar tetra-quark resonance as $f_0(ud\bar u\bar d)$ 
by treating only the connected diagram in quenched anisotropic lattice QCD.
Note that the tetra-quark D$_{\rm s0}^{++} (cu\bar s\bar d)$ 
corresponds to the connected diagram of $f_0(u\bar ud\bar d)$ 
in the unrealistic idealized SU(4)$_f$ limit of $m_{u,d}$=$m_s$=$m_c$. 
Then, the lattice QCD study of the connected diagram of $f_0(u\bar ud\bar d)$ 
means that of the manifestly exotic tetra-quark D$_{\rm s0}^{++} (cu\bar s\bar d)$ 
with $C=+1$, $S=+1$, $I_3=+1$ in the idealized SU(4)$_f$ limit.\cite{STIO06}

\section{Anisotropic Lattice QCD and Hybrid Boundary Condition Method}

To examine exotic 4Q states as $f_0(ud\bar u\bar d)$, 
we adopt a diquark-type interpolating field,
$
  O
  \equiv
  \epsilon_{abc} \left(u_a^T C\gamma_5 d_b \right)
  \times
  \epsilon_{dec} \left(\bar u_d C\gamma_5 \bar d^T_e \right),
$
which is expected to have a small overlap with  
two-meson scattering states.
In lattice QCD, we calculate and analyze the temporal 4Q correlator 
$G(t)\equiv \frac{1}{V}\sum_{\vec x}\langle O(t,\vec x)
O^\dagger (0,\vec 0)\rangle$, 
where the total momentum of the 4Q system is projected to be zero.
We impose the Dirichlet boundary condition for quarks 
in the temporal direction to avoid the backward meson propagation.\cite{TUOK05}

To get detailed information on the temporal behavior of the 4Q correlator $G(t)$, 
we adopt anisotropic lattice QCD 
with the plaquette action and the $O(a)$-improved Wilson 
fermion at $\beta \equiv 2N_c/g^2=5.75$ on $12^3 \times 96$ with renormalized anisotropy $a_s/a_t=4$. 
We use 500-1000 gauge configurations.
The temporal lattice spacing is $a_t\simeq$ 0.045fm, and the spatial lattice size is $L =12 a_s \simeq 2.15{\rm fm}$. 
We take $\kappa$=0.1210, 0.1220, 0.1230 and 0.1240, 
which cover the quark mass region of $m_s< m_q <2 m_s$.

To examine low-lying 4Q resonances, 
we use the ``Hybrid Boundary Condition" (HBC) \cite{IDS05,STIO06,IDIST06}, 
where we impose the anti-periodic boundary condition (APBC) for quarks ($u$,$d$) and 
the periodic boundary condition (PBC) for antiquarks ($\bar u, \bar d$), as shown in Table~I.
By applying HBC on a finite lattice with $L^3$, the two-meson threshold is raised up, 
while the mass of a compact 4Q resonance is almost unchanged. 
Then, the 4Q resonance may become visible as a low-lying state in HBC, if it exists.

For a compact $qq\bar q\bar q$  resonance, 
since it contains even number of quarks, it obeys PBC even in HBC, 
and therefore its energy in HBC is almost the same as that in PBC \cite{STIO06}.
For a two-meson scattering state, 
both mesons obey APBC and have non-zero relative momentum 
$\vec p_{\min}=(\pm\frac{\pi}{L},\pm\frac{\pi}{L},\pm\frac{\pi}{L})$, {\it i.e.},
$|\vec p_{\rm min}|=\sqrt{3}\pi/L$ in HBC.
In fact, the two-meson threshold $m_1+m_2$ ($m_1$, $m_2$: meson masses) in PBC 
is raised up in HBC as 
$E_{\rm th} \simeq \sum_{k=1,2}\sqrt{m_k^2+\vec p_{\rm min}^2}$ 
with $|\vec p_{\rm min}| =\sqrt{3} \pi/L \simeq 0.5 {\rm GeV}$ for 
$L \simeq 2.15 {\rm fm}$.

\begin{table}[h]
\newcommand{\cc}[1]{\multicolumn{1}{c}{#1}}
\caption{
Hybrid Boundary Condition (HBC) to raise 
the threshold of two-meson scattering states.}
\begin{center}
{\large
\begin{tabular}{llllll} \hline \hline
  & $u$, $d$ & $\bar u$, $\bar d$  & $q\bar q$-meson & two-meson threshold & 
tetra-quark ($qq\bar q\bar q$)\\
\hline 
PBC & periodic & periodic & periodic & $m_1+m_2$ & ~~~~~periodic \\
HBC & anti-periodic & periodic & anti-periodic & 
$\sum_{k=1,2}\sqrt{m_k^2+\vec p_{\rm min}^2}$ & ~~~~~periodic  \\
\hline\hline
\end{tabular}
}
\end{center}
\end{table}

\section{Effective-Mass Analysis for Four-Quark Systems}

We perform the standard effective-mass analysis for the low-lying 4Q state \cite{STIO06}.

For comparison, we first calculate the ordinary $q\bar q$ scalar-meson mass 
from the connected diagram in anisotropic quenched lattice QCD. 
We show in Fig.1(a) the lowest $q\bar q$ scalar meson mass plotted against $m_\pi^2$.
Our quenched lattice QCD indicates that the lightest $q\bar q$ scalar meson 
has a large mass as $M_{q\bar q} \simeq 1.310(55)$GeV after the chiral extrapolation. 
Thus, the lightest $q\bar q$ scalar meson is identified with 
the multiplet of $f_0(1370)$ and $a_0(1450)$ \cite{STIO06}, which 
is consistent with the quark-model assignment for scalar mesons.
(If disconnected diagrams are dropped off, 
isoscalar and isovector $q\bar q$ mesons degenerate in quenched QCD, 
{\it e.g.}, $m_\rho=m_\omega$, $m_{f_0}=m_{a_0}$.)

\begin{figure}[h]
\begin{center}
\begin{minipage}{4.5cm}
\includegraphics[width=3.5cm,angle=270]{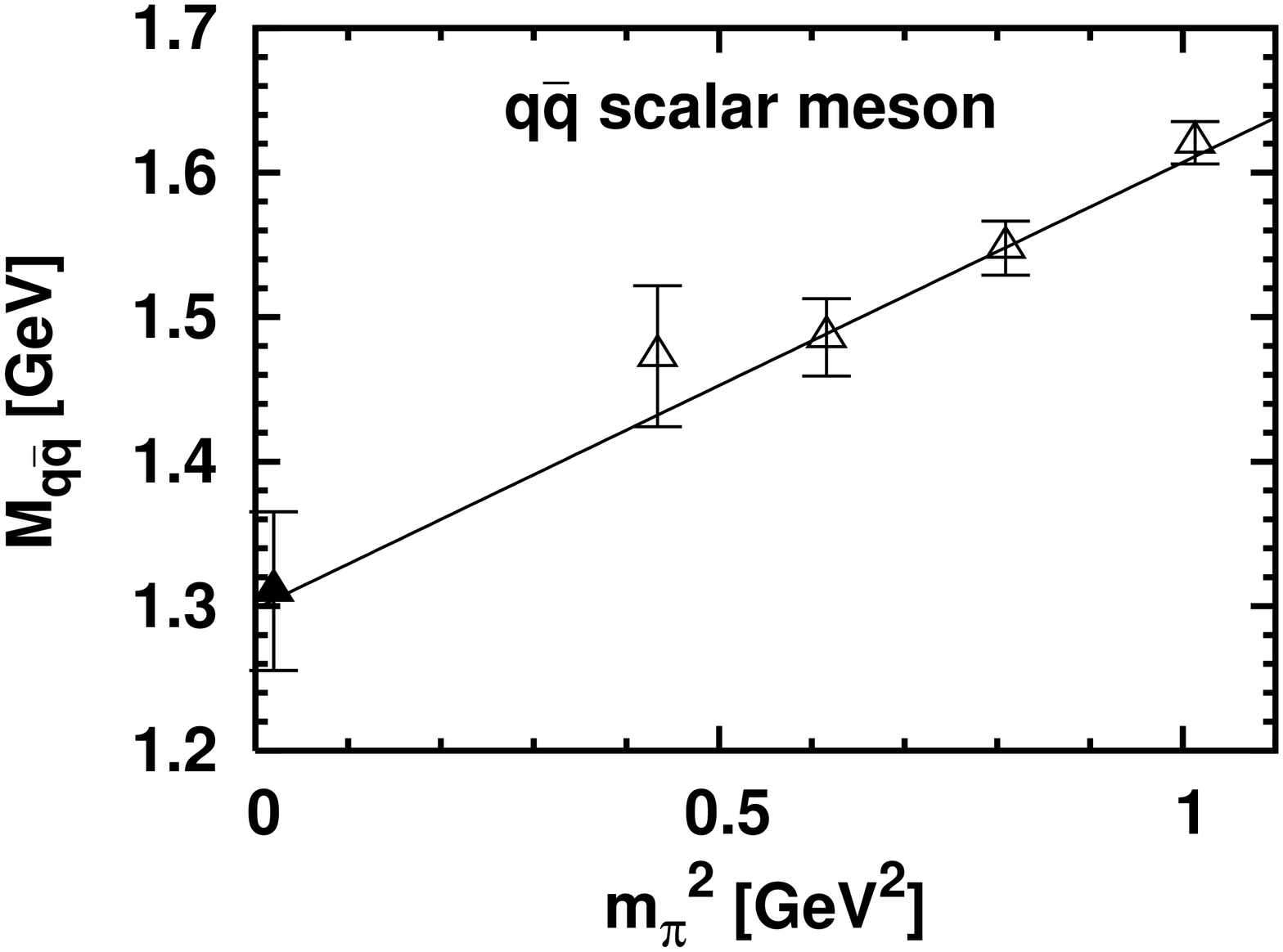}
\vspace{-0.1cm}
%
\label{fig1}
\end{minipage}
\hspace{-0.3cm}
\begin{minipage}{4.5cm}
\includegraphics[width=3.5cm,angle=270]{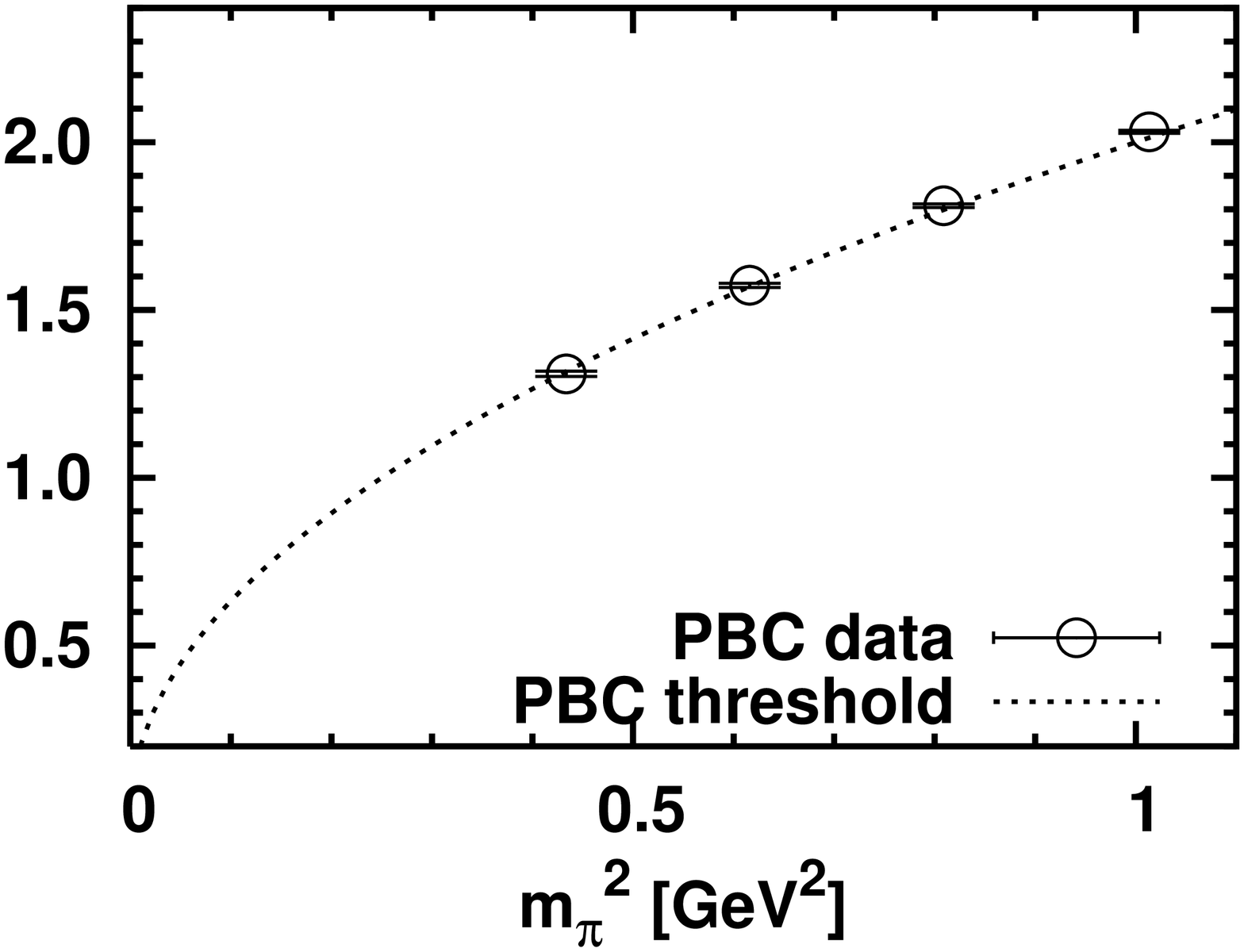}
\end{minipage}
\begin{minipage}{4.5cm}
\includegraphics[width=3.5cm,angle=270]{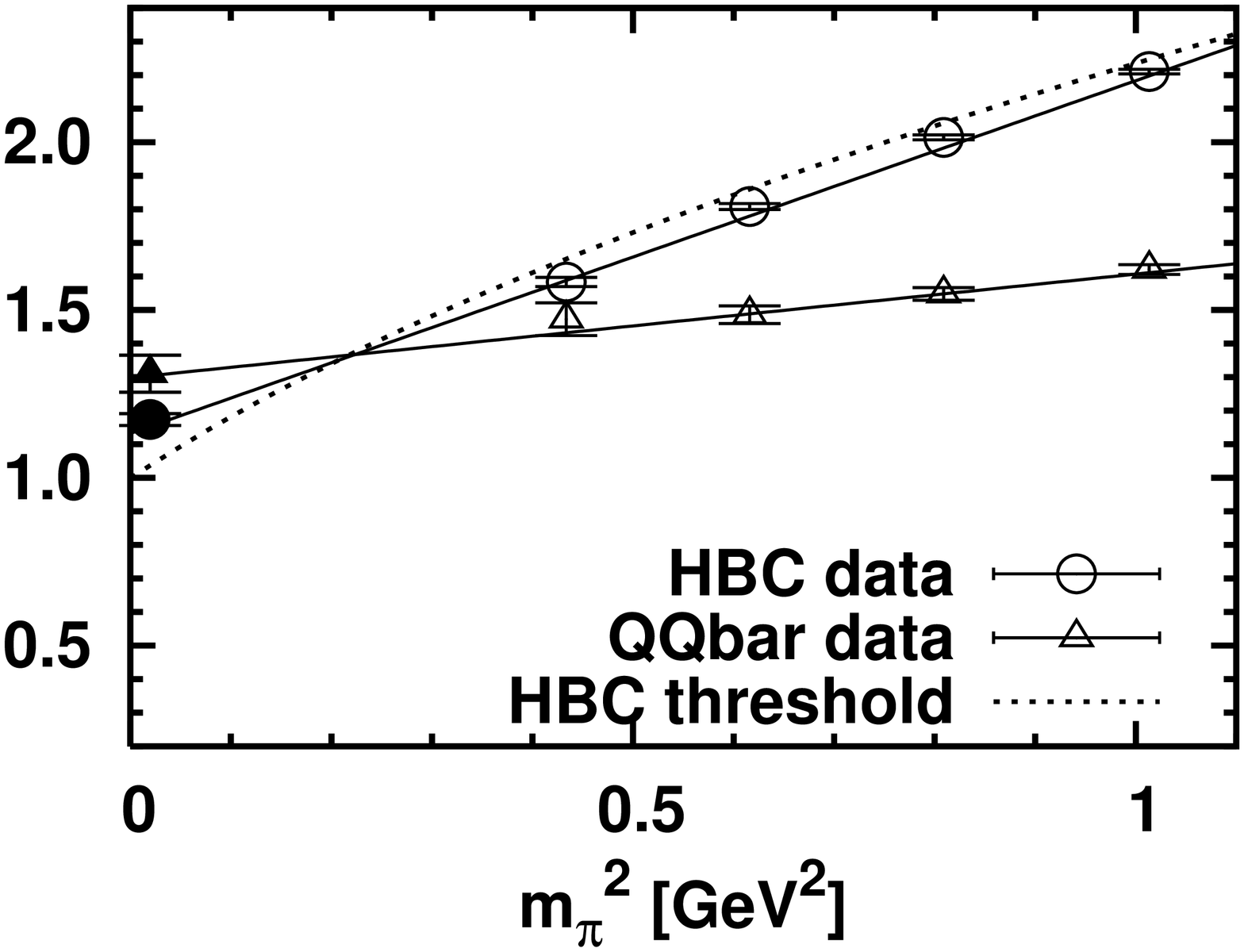}
\end{minipage}
\caption{
(a) The lowest mass $M_{q\bar q}$ of the $q\bar q$ scalar meson plotted against $m_\pi^2$.
The $q\bar q$ scalar meson mass $M_{q\bar q}$ is extracted from the connected diagram 
in quenched lattice QCD.
(b) 
The lowest mass $M_{\rm 4Q}$ of the scalar 4Q ($ud\bar u\bar d$) system 
against $m_{\pi}^2$ in PBC.
The curve denotes the two-pion threshold.
(c) 
The lowest mass $M_{\rm 4Q}$ of the scalar 4Q system against $m_{\pi}^2$ in HBC.
Open (closed) symbols are the direct lattice QCD data (chiral extrapolated values).
The curve denotes the raised two-pion threshold $E_{\rm th}$ in HBC.
Triangles denote the $q\bar{q}$ scalar-meson mass for comparison.
}
\end{center}
\vspace{-0.1cm}
\end{figure}

We calculate the lowest energy of the scalar 4Q ($ud \bar u \bar d$) system 
in PBC and in HBC, as shown in Figs.1 (b) and (c).
In PBC, the lattice  QCD data  coincide with the two-pion threshold curve, 
and hence the lowest 4Q state is just a two-pion scattering state.
In HBC, the 4Q state appears slightly below the raised two-pion threshold $E_{\rm th}$ 
by 40-75MeV and seems to behave linearly in $m_{\pi}^2$ rather than the two-pion threshold.
Then, there are two possible interpretations for the lowest 4Q state in HBC: 
i) If the slight difference from the two-pion threshold can be explained by 
the two-pion interaction and the finite-box effect in HBC, 
the lowest 4Q state in HBC is also regarded as a two-pion scattering state;
ii) From the slight difference from the two-pion threshold and different chiral behavior, 
one may expect that the 4Q state in HBC differs from the two-pion scattering state.  
If a linear chiral extrapolation is adopted, the 4Q mass is estimated as 
$m_{\rm 4Q} \simeq 1.174(18)$GeV in the chiral limit.\cite{STIO06}
 
 \section{MEM Analysis for Four-Quark Systems}

 \begin{figure}[ht]
  \begin{center}
   \begin{minipage}{0.45\linewidth}
    \includegraphics[width=\linewidth]{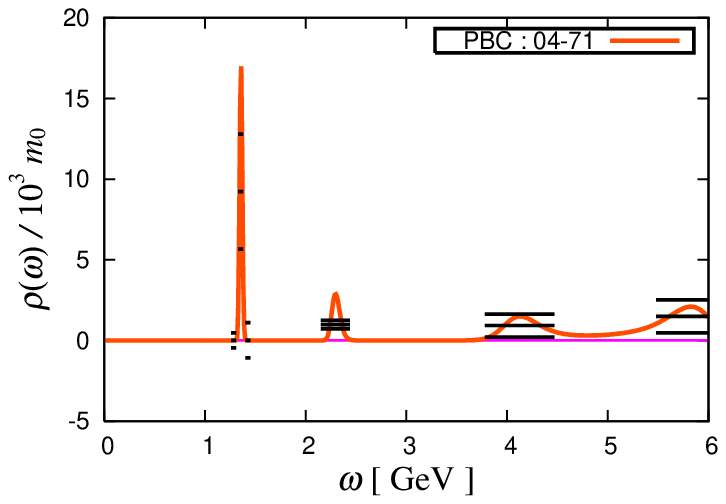}
   \end{minipage}
   \begin{minipage}{0.45\linewidth}
    \includegraphics[width=\linewidth]{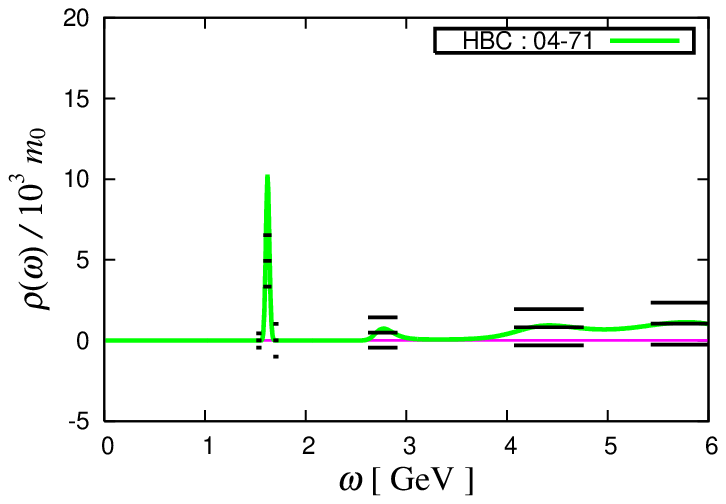}
   \end{minipage}
  \end{center}
\vspace{-0.1cm} 
  \caption{
The dimensionless spectral function $\rho(\omega)\equiv A(\omega)/\omega^8$ 
obtained from the point-source point-sink 4Q correlator $G(t)$
for $\kappa$=0.1240 ({\it i.e.}, $m_q \sim m_s$) in PBC (left) and HBC (right).
  }
  \label{fig:errorbar_PBC_HBC}
\vspace{-0.25cm} 
\end{figure}
 
To clarify whether the 4Q resonance exists or not, 
we perform the Maximum Entropy Method (MEM) analysis for the scalar 4Q system 
in PBC and in HBC, 
and obtain the spectral function $A(\omega)$ of the scalar 4Q state 
from the 4Q correlator $G(t)$ with $\kappa$=0.1240, as shown in Fig.2.
Here, we calculate the default function $m(\omega)$ of the 4Q system 
with the lowest-order perturbative QCD:
$
m(\omega)
=\frac{4N_c}{2^8 \Gamma(5)\Gamma(6)\pi^6}\omega^8=m_0\omega^8,
$
which gives the asymptotic form of $A(\omega)$ in the high-energy region.
 
In both cases of PBC and HBC, the lowest peak around 1.5GeV exhibits a large strength.
Note however that, in the finite-size lattice, 
the momentum of the pion is discretized, and all the scattering states are observed as 
resonance-like states in lattice QCD.
By changing the spatial boundary condition from PBC to HBC, 
all the low-lying peaks in the spectral function are significantly shifted to the high-energy side.
Since each low-lying peak is strongly affected by the spatial boundary condition,
it is not a spatially-localized resonance 
but a spatially-extended two-pion scattering state, {\it i.e.}, a discretized continuum state.
(If there is a low-lying pole whose position is unchanged between PBC and HBC, 
it is to be a spatially-localized resonance.\cite{IDIST06})

Thus, from the combination analysis of MEM with HBC, 
we finally conclude that the 4Q system appears as a two-pion scattering state 
and there is no spatially-localized 4Q resonance 
in the quark-mass region of $m_s< m_q <2m_s$.

\vspace{-0.1cm}

\section*{Acknowledgements}
H.S thanks Profs. K.-F.~Liu and K.~Terasaki for useful discussions. 
The authors are thankful to the Yukawa 
Institute for Theoretical Physics at Kyoto University.
Discussions during the YKIS2006 on ``New Frontiers QCD" were useful to complete this work. 
The lattice QCD calculations were done on 
NEC-SX5 at Osaka University.

\end{document}